\documentstyle[aps]{revtex}

\title{Fractional-differential equations of motion}

\author{Alexander I. Olemskoi}
\address{Department of Physical Electronics,
Sumy State University\\ 2, Rimskii-Korsakov St., 244007 Sumy,
UKRAINE\\
E--mail: Alexander@olem.sumy.ua}

\begin{document}

\maketitle

\begin{abstract}
An elementary system leading to
the notions of fractional integrals and derivatives is considered.
Various physical situations whose description is
associated with fractional differential equations of motion are discussed.



\end{abstract}

\vspace{0.3cm}

Consider a medium exhibiting memory. Such a situation arises in
describing the structure of a solid far from thermodynamic
equilibrium: in amorphous materials, under structural relaxation of
high-$T_c$ oxide superconductors,
under plastic deformation and destruction of solids, in
solid solutions and in martensite macrostructures, and so on \cite{1}. The
existence of memory implies that if a force $f(t')$ acts on a system at an
instant $t' > 0$, then a flux $J$ arises whose magnitude at a subsequent
instant $t > t'$ is given by
\begin{equation}
J(t)=\int_0^t M(t-t')f(t')\mbox{d}t'.
\end{equation}
For a system without memory, the time dependence of the memory
function $M(t-t')$ has the form
\begin{equation}
M(t-t')=\gamma \delta (t-t'),
\end{equation}
where $\gamma$ is a positive constant and
$\delta(t-t')$ is the Dirac
$\delta$-function. On substituting (2) into (1), we obtain the relation
\begin{equation}
J(t)=\gamma f(t),
\end{equation}
according to which in the absence of memory only
a force value at the same instant $t$ has an
effect on the flux $J(t)$. With memory inclusion, the
$\delta$-function in (2) spreads into a bell-shaped function with
a width determining an interval $\tau$ during which the force $f$ has an
effect on the flux $J$. For systems with ideal memory, we have
$\tau\to\infty$,
i.e., the flux $J(t)$ is formed over all the course of the action
of the force $f(t')$ up to the instant $t$. Formally, this is expressed
by specifying the kernel of the integral relation (1) in the
form
\begin{equation}
M(t-t')=\gamma /t.
\end{equation}
Here, the dependence on the instant $t'$ of the force $f$ action is
absent, and the dependence of the memory on the flux measurement time $t$
is given so as to satisfy the normalization condition
\begin{equation}
\int_0^t M(t-t')\mbox{d}t'=\gamma.
\end{equation}

Relation (1) as written in the time domain, is inconvenient because
of the convolution (integral over $t'$). This can be eliminated by
using the Laplace transformation
\begin{equation}
f(t)=\int^{+i\infty}_{-i\infty}f(\lambda
)\mbox{e}^{-\lambda t}{\mbox{d}\lambda \over 2\pi i},\qquad
f(\lambda)=\int_{0}^{\infty}f(t)\mbox{e}^{\lambda t}\mbox{d}t,
\end{equation}
which enables one to move from the time $t$ to the complex frequency
$\lambda$. After transforming both sides of the definition (1), this
converts to the algebraic form
\begin{equation}
J(\lambda )=M(\lambda)f(\lambda ).
\end{equation}
Obviously the Laplace transform of the kernel (2), which
corresponds to the absence of memory, reduces to a constant:
\begin{equation}
M(\lambda )=\gamma .
\end{equation}
For ideal memory, we obtain from (4), in the limit $|\lambda |t\gg 1$
\begin{equation}
M(\lambda )=\gamma /\lambda t.
\end{equation}
Thus, the inclusion of memory leads to the transformation of the
constant kernel (8) into the hyperbolic form (9).

In this
connection, one could ask: How does the function
$M(\lambda)$ stand up if memory is complete but nonideal? This means
that the memory manifests itself within the interval $(0,t)$ preceding
$t$ but not at all instants $t'$. Assume, for example, that memory
holds only at the points of a Cantor set. One can then
expect that its fractal dimension $D$ will be associated with a measure
of memory preservation.

Although elementary for simple time dependences (2) and (4), the
problem of finding the Laplace transform for the memory acting at the
points of the Cantor set is much more difficult.
The appropriate calculations \cite{2} arrive at the result
\begin{equation}
M(\lambda )=A_Dz^{-D}, \qquad z=(1-\xi )\lambda t.
\end{equation}
Here, the fractal dimension $D$ is defined by usual equality
\cite{1} $D=\ln j/\ln \xi^{-1}$, $j$ is the branching index
of the corresponding hierarchical tree, $\xi$ is
the similarity parameter, and the constant $A_D=2^{-D/2}$ for the Cantor
discontinuum with the number of elementary blocks $j = 2$.

It can easily be seen that the memory function (10) satisfies the
similarity condition $M(\xi\lambda )=\xi^{-D}M(\lambda )$
with the exponent $D$. Being the fractal
dimension of the Cantor set at whose points memory is switched on,
this exponent thus represents a quantitative measure of the
manifestation of memory effects. For an empty Cantor set ($\xi =0$),
one has $D = 0$, and the dependence (10) reduces to
the constant (8), corresponding to the entire absence of memory.
The exponent $D$ increases with the similarity parameter $\xi > 0$,
and the Laplace transform (10) becomes an increasingly more rapidly varying
function. The limiting value $\xi = j^{-1}$ of the similarity parameter
yields the dimension $D = 1$, which corresponds to the ideal memory
governed by the function (9).

Thus, systems with residual memory are described by the Laplace
transform (10), where the exponent value $0\le D\le1$
determines the extent of memory
preservation. Using the inverse Laplace transformation (6), we
find for the time-dependent memory function in the simplest
case of the Cantor set with $j = 2$ (see \cite{1}, \cite{2})
\begin{equation}
M(t-t')=(\gamma /\pi )\left[ \sqrt{2} (1-\xi )t\right] ^{-D}\Gamma
(1-D) (t-t')^{D-1},
\end{equation}
where $\Gamma(1-D)$ is the gamma-function. Substituting this into
equality (1), we bring this equality in the form
\begin{equation}
J(t)=(\gamma /\pi )\left[ \sqrt{2} (1-\xi)\right] ^{-D}\widehat{D}^{-D}f(t),
\end{equation}
where the integral is introduced
\begin{eqnarray}
\widehat{D}^{-D}f(t)&=&\Gamma (1-D)\int_0^1
(1-u)^{D-1}f(ut)\mbox{d}u\nonumber\\
&=&D^{-1}\Gamma (1-D)\int_0^1 f\left(
(1-u)t\right) \mbox{d}u^D,
\end{eqnarray}
whose fractional nature is reflected by the presence of the
exponent $D$ in the argument $u$ of differential \cite{3}.

In the foregoing discussion we have used an integral representation for
memory effects. It is easy to show, however, that an equivalent
differential equation can be written that corresponds to this
integral representation. For example, if we have some conserved
quantity $n$ (like the number density of atoms),
then it space--time dependence $n({\bf r},t)$ is governed by the continuity
equation
\begin{equation}
{\partial n\over \partial
t}=-\nabla {\bf J},
\end{equation}
where the flux {\bf J} is given by equality (1). In the absence of
memory effects, equality (3) is satisfied, where the force {\bf f} is
given by
\begin{equation}
{\bf f}=-\nabla \mu,
\end{equation}
where $\mu$ is the chemical potential, and (14) leads to a
conventional equation of diffusion type \cite{4}
\begin{equation}
{\partial n\over \partial t}=\nabla (\gamma\nabla \mu).
\end{equation}
According to the earlier discussion, when memory is switched on, a
fractional integral appears on the right-hand side of (16). The treatment of
the resulting integro-differential equation, with partial derivatives
and furthermore, of fractional order, presents a very difficult
problem.

However, this problem can be simplified when it is considered that the
memory function $M(t-t')$ vanishes within some intervals, thus leading
to a fixing of the flux {\bf J} in (14) and correspondingly to the
vanishing of the
velocity $\partial n/\partial t$.
Therefore, it would appear natural that loss of
memory can be taken into account by a fractional property, not only of
the flux integral (1) but also of the time derivative in (14).
Because equation (16) corresponds to the entire absence of memory
($D = 0$), when it is switched on, as is reflected by the growth of the
exponent $D$, the rate of $n$ variation must decrease, and it is reasonable
to postulate a fractional-differential equation of the form (see further)
\begin{equation}
{\partial ^ q n\over \partial
t^ q}=\nabla (\gamma\nabla \mu), \qquad  q =1-D.
\end{equation}
As would be expected, the growth of the memory parameter $D$ leads to
lowering the order $ q$ of this equation. In systems with ideal
memory ($D = 1$), the derivative on the left-hand side of (17)
disappears so that one should treat the behavior of the flux (1)
itself which is what we have done above.

Our considerations in favor of the specific form $ q = 1 -D$
for the fractional degree in equation (17) act as guidelines. To
confirm these, we will show that equation (12) for the flux in the
form of the fractional integral (13) is in agreement with the
solution of the fractional-order differential equation (17). Indeed,
replacing the variable $u$ by $(1-u)t$ in the integrand of the second
integral of (13), one can easily show the time dependence of the
flux to have the form $J\propto t^{-D}$.
Substituting this dependence into (14), we
find the solution
\begin{equation}
n\propto t^{1-D},
\end{equation}
satisfying the tested equation (17). This confirms conclusively the
equivalence of the mutually complementary concepts of the
fractional integral (13) and the fractional-differential equation
(17). Notably is displayed in
the same form the geometric relation $L\propto l^{1-D}$
for coastline length $L(l)$
and in the physical dependence $n(t)$
obtained. In particular, it follows that the time $t$ plays the some role as
the length of the segments $l$,
covering a $D$-dimensional fractal set in the space
of states, the points of this set determining the memory of the system.

When interpreting equation (17) in fractional derivatives,
we start from the fact
that in the usual case $D = 0$ this equation governs irreversible
processes like diffusion, heat conductivity, and so forth \cite{4}.
In the latter processes, the microscopic
memory with respect to time inversion is completely lost:
the mechanical equations of
motion for an isolated object exhibit ideal memory, manifesting itself
in the invariance with respect to the
replacement $t\to -t$, whereas such invariance is
completely violated in the thermodynamic equation (16). It follows that
a decrease in the derivative order $ q = 1 - D$ in (17)
corresponds to the switching on of memory channels whose part is
determined by the fractal dimension $D$. The remaining
$ q = 1 - D$
channels provide the irreversibility of the system. In the initial
form, the residual irreversibility can be taken into account by using
the fractional form of the Liouville equation \cite{2}
\begin{equation}
{i\hbar \over t}{\partial ^D\rho \over \partial
u^D}=\left[ H,\rho \right] ,
\end{equation}
where $\hbar$ is the Planck constant, $H$ is the Hamiltonian, $\rho$ is the
nonequilibrium statistical operator, $t$ is the evolution time of a
macroscopic system, and $u = t'/t$ is the dimensionless microscopic
time bounded by the condition $u < 1$. It is apparent that the
condition
$D < 1$ enables one, in the framework of (19), to take into account the
irreversible effects in the interval $(0,t)$ which are due to the loss
of $ q = 1 - D$ deterministic channels. Such a means of allowing
for the irreversibility is obviously preferable to the
phenomenological addition to the right-hand side of (19) of the
relaxation term $-\rho/\tau$, where $\tau$ is the relaxation time,
\cite{4} or to the inclusion of an infinitesimal source \cite{5}).

Apart from equations of the diffusion type, one might expect the
appearance of fractional derivatives in a case
of the description of the motion of
particles under inelastic collisions. Performing calculations similar
to those that led to the flux of the form (12), one can show
\cite{2} that, if a force $F$ acts on a particle of mass $m$
in each collision, then the change in its velocity
\begin{equation}
\Delta v=\left[ \sqrt{2}(1-\xi )\right]
^{-D}(t/m)\widehat{D}^{-D}F
\end{equation}
is determined by a fractional integral of the form (13). From the
above example of the diffusion equation, one can see that, in order to
switch to the corresponding fractional-differential equation, it is
sufficient to operate on equation (20) with the operator
$\widehat{D}^D=\partial ^D/\partial u^D$ which is
inverse with respect to the fractional integral $\widehat{D}^{-D}$.
In the case of an
elastic force ${\bf F}=-\lambda \Omega \nabla ^2{\bf r}$,
where $\lambda$ is the elastic constant and
$\Omega$ is the atomic volume, we then obtain the generalized transport
equation for the particle coordinate ${\bf r}$ \cite{2}
\begin{equation}
\frac{{\rm d}^{1+D}{\bf r}}{{\rm d}u^{1+D}}+(ct)^2\nabla ^2{\bf r}=0,
\end{equation}
where the dimensionless time $u = t'/t < 1$ is used and the characteristic
velocity $c$, defined by
\begin{equation}
c^2=(\lambda
\Omega /m)\left[ \sqrt{2}(1-\xi )\right] ^{-D}.
\end{equation}
is introduced. Equation (21) governs a new type of wave motion that is intermediate
between conventional diffusion ($D = 0$) and classical waves
($D = 1$). Accordingly, formula (22) determines the diffusion
coefficient $D = c^2t$ in the first case and the wave velocity $c$ in
the second case.

The fractional-differential equations (17) and (21) are written
as applied to the study of the space--time behavior of conserved
quantities. As we known \cite{6}, this
is reflected in the presence of a second derivative with respect to the
coordinate in the right-hand sides of these equations. For a
nonconserved quantity
$\eta$, this derivative vanishes, and the
diffusion equation (17) converts to the generalized Landau--Khalatnikov
relaxation equation
\begin{equation} {{\rm d}^{1-D}\eta \over {\rm d}u^{1-D}}=
-(\gamma_Dt)\eta ,
\end{equation}
and the wave equation (21) converts to the generalized equation for an harmonic
oscillator
\begin{equation}
{{\rm d}^{1+D}\eta \over {\rm d}u^{1+D}}+(\omega_Dt)^2\eta =0;
\end{equation}
here, as above, $u = t'/t$ is the dimensionless time, and the parameters
$\gamma_D$, $\omega_D\propto \left[
\sqrt{2}(1-\xi)\right]^{-D}$ determine the relaxation time
$\tau_D=\gamma^{-1}_D$ and
the oscillator frequency $\omega_D$. The fractal dimension $D$ specifies a
measure of the residual memory of the system. In its absence ($D = 0$), both
equations (23) and (24) reduce to the equation of Debye's
relaxation which governs the behavior of the simplest thermodynamic
systems ($\gamma_0=\omega_0^2t$ in this case).
For ideal memory ($D = 1$),
equation (23) degenerates into the condition of phase equilibrium
$\eta = \eta_0$ and (24) degenerates into the ordinary
oscillator relation.

The above consideration shows that, in limiting cases, one can recognize
the following equations of motion, which correspond to the different types
of behavior of a continuous medium. First, there is the continuity
equation, being the condition for the conservation of particles in the
medium. In the context of a mechanistic approach, the oscillatory
behavior is governed either by an equation of oscillation, which
involves a second derivative with respect to time, or by a wave
equation involving second derivatives with respect to both time
and space coordinates. Under the thermodynamic
description of a system determined by a hydrodynamic mode, its amplitude
(order parameter) may be characterized both by reactive and by
dissipative regimes. In the first regime, the equation of motion
reduces to an oscillatory equation for a nonconserved parameter and
to a wave equation for a conserved one. Upon entering the
dissipative regime, the time-derivative order is lowered to the first
order.

Although these equations govern quite different physical situations,
our discussion shows them to transform smoothly from one to another
\cite{2}, \cite{7}. Thus, the
question arises: Is it possible to give a model of the medium whose
equation of motion includes all the above types as special cases?
Following \cite{8}, we will show how such an
equation can be obtained.

We start from the continuity equation
$\dot{\eta}+\nabla {\bf j}=0$
for the order parameter field $\eta({\bf r},t)$ characterized by the
flux ${\bf j}({\bf r},t)$.
In the general case, the latter is nonlocally related to the
distribution of the chemical potential $\mu({\bf r},t)$.
In the framework of
the Fourier representation with respect to the space coordinate and of the
Laplace representation with respect to time (we take into account the causality
condition), this relation reduces to the equality
${\bf j}({\bf k},\lambda )={\bf M}({\bf k},\lambda )\mu
({\bf k},\lambda )$, where ${\bf k}$ is
the wave vector and $\lambda$ is the complex frequency.

To find the dependence of the memory function ${\bf M}({\bf k},\lambda)$
on ${\bf k}$,
we will consider the simplest one-dimensional model of the medium in
the form of a segment $0\le x\le l$.
Then the flux through the right boundary
is given by (cf.Eq.(1))
\begin{equation}
j(l)=\int_0^lM(l-x)\mu
(x){\rm d}x.
\end{equation}
We assume that the function $M(x)$ takes on nonzero values only at the
points of the Cantor set which is obtained from the original
segment $[0,l]$ by the convolution method. Then, as in Ref.\cite{2},
it can be shown that the desired dependence is given by the
equality similar to (10):
\begin{equation}
M(k)={\rm e}^{-\alpha D}\kappa ^{-D},\qquad \kappa =ikl(1-\xi ).
\end{equation}
Here, the constant $\alpha\sim1$ is introduced (according to \cite{2},
$\alpha =2^{-1}\ln 2=0.347$ for the simplest Cantor set),
$D$, $\xi$,
and $l$ are the fractal dimension, the similarity parameter, and
the scale, respectively, which describe the set of points on
the $x$-axis where the order parameters conserved. Using (26), we find
the dispersion law from the continuity equation:
$$
\lambda =
B_{-D}k^{1-D}, \qquad B_{-D}=i^{1-D}{\rm e}^{-\alpha D}(1-\xi
)^{-D}l^{1-D}{\rm d}\mu /{\rm d}\eta .\eqno(27a)
$$
When we reduce the fractal set to a continuum ($\xi = 1/2$), its
dimension is then $D = 1$, and relation (27a) corresponds to the
Landau--Khalatnikov equation for the nonconserved order parameter. In
the general case of $\xi < 1/2$, upon reduction of the fractal dimension of
the set with ordering points, the vanishing of long-wave harmonics
occurs, which is inherent in the order parameter conserved. However,
even the maximum value of the exponent $1-D$, corresponding to the empty
set ($\xi = 0$), proves to be half as large as the exponent 2 inherent
in the order parameter conserved.

To raise the exponent in the
dispersion law (27a), we assume that the order parameter becomes
conserved after going to the empty set as a result of the reducing the
similarity parameter $\xi$ from $1/2$ to 0. Next, using a procedure
inverse to convolution, we increase the fractal dimension to a
finite value $D > 0$. Obviously, this can be done by decreasing the
similarity parameter $\xi$ from $\infty$ to a finite value $\xi\ge2$. The
condition $\xi > 1$ implies that we stretch the segments rather than
convolute them. Then, the functional equation $
M(\kappa /\xi )=2^{-1}M(\kappa )$
takes the form $M(\xi\kappa )=2M(\kappa )$. Its
solution $M(\kappa )={\rm e}^{\alpha D}\kappa^D$ has a positive
exponent whose value is now determined by the relation
\setcounter{equation}{27}
\begin{equation}
D=\ln2/\ln \xi
\end{equation}
instead of the usual equation $D=-\ln2/\ln \xi$.
Accordingly, the dispersion law for the partially conserved order
parameter has the form
$$
\lambda = B_Dk^{1+D},
\qquad B_D=i^{1+D}{\rm e}^{\alpha D}|1-\xi |^Dl^{1+D}{\rm d}\mu /
{\rm d}\eta .\eqno(27)
$$

As seen from (27a), the distinction from the case of the conserved
order parameter reduces to the change of signs of $D$ and $\xi - 1$.
According to (28), this corresponds to the transition from
contraction ($\xi < 1$) to stretching ($\xi > 1$) when constructing
a fractal.

Thus, if it is assumed that the similarity parameter $\xi$ is defined
in the regions $(0,1/2]$ and $[2,\infty)$,
the dispersion law will encompass
both the cases of the nonconserved ($D < 0$) and conserved ($D > 0$) order
parameter. The transition between the limiting cases occurs due to a
smooth variation in the similarity parameter $\xi$. At $\xi = 1/2$,
there is $D= -1$, and we have complete nonconservation of the order parameter.
A decrease in $\xi$ down to values $0 < \xi < 1/2$ partially violates the
nonconservation conditions, thus leading to an increase in the fractal
dimension in the range $-1 < D < 0$ and to the rise in the exponent of
the dispersion law (27). With the empty Cantor set ($D = 0$),
we go from the nonconserved order parameter ($\xi\to0$) to the
conserved one ($\xi\to\infty$). A decrease in the similarity parameter
within the segment $2 < \xi < \infty$ ($0 < D < 1$) provides the rise in the
number of conservation channels, thus increasing the exponent of the
dispersion law (27). Finally, with the complete conservation of
the order parameter ($\xi = 2$, $D = 1$), we obtain, as expected,
$\lambda\propto k^2$.


It is easy to understand that this variation in the order of the
derivative with respect to the space coordinate can be extended to the
derivative with respect to time. Indeed, as shown in \cite{2}, the
memory switching on relative to the time inversion $t\to-t$ in the equation of
particle motion, leads to the appearance in (27) of the exponent
$1-D_0$ at frequency $\lambda$, where $D_0$ is the dimension of a fractal
set on the time axis, at the points $t$ of which the mechanical memory is
switched on. The set itself is represented by the similarity
transformation $F(z/\xi)=2^{-1}F(z)$,
where $z = \lambda t(1-\xi_0)$ is the argument
and $0\le\xi_0\le1/2$.
The transition from the contraction region $\xi_0\le1/2$ into
the stretching region $\xi_0\ge2$, corresponding to the functional equation
$M(\xi_0 z)=2M(z)$,
occurs on the empty set ($D_0 = 0$) and implies the replacement
of the corpuscular memory by the wave memory. The latter is
enhanced with a decrease in the similarity parameter $\xi_0$ and
becomes ideal at $\xi_0 = 2$. Then, the exponent at the frequency $\lambda$
in (27) takes on the value 2 inherent in the wave equation.


Thus, in the general case of partial nonconservation/conservation of
the order parameter and of the non-ideal memory of particle/wave, the
dispersion law for a one-dimensional fractal medium takes the form
\begin{eqnarray}
\lambda &=&c^{1/(1+D_0)}(ik)^ q;\nonumber\\
 q &=&(1+D)/(1+D_0), \nonumber\\
c&=&{{\rm e}^{\alpha D}|\xi-1|^Dl^{1+D}\over {\rm e}^{\alpha _0D_0}
|\xi_0-1|^{D_0}\tau^{1+D_0}}{{\rm d}\mu \over {\rm d}\eta}, \qquad
\alpha_0, \alpha \sim 1,
\end{eqnarray}
where $l$ and $\tau$ are the characteristic space-time scales, and the
dependence of the dimension $D_0$ upon the parameter $\xi_0$ is
determined by the same equality (28) as for the space component
$D(\xi)$. Accordingly, the fractional-differential equations of motion
are written in the form\footnote{In the $d$-dimensional case, the second
term of equation (30) is replaced by a sum of terms characterized by
scale $l_i$, fractional dimension $D_i$, similarity parameter $\xi_i$ and
constant $\alpha_i$, for $i=1,\ldots, d$.}
\begin{equation}
{{\rm d}^{1+D_0}\eta \over {\rm d}t^{1+D_0}}+
i^{1+D}{{\rm d}^{1+D}(c\eta )\over {\rm d}x^{1+D}}=0.
\end{equation}
In the linear approximation $\mu \propto \eta$, the constant $c$
may be factored out of the differentiation sign.

It can be easily seen that, in the limiting cases of the
integer-valued space-time dimensions $D_0$ and $D$, the linearized
equation (30) encompasses the main regimes of medium evolution. For
example, in the case $D_0 = 1$, this governs the wave motion at $D = 1$,
uniform oscillations at $D = -1$, and localized oscillations at $D = 0$.
The constant $c$ equals the square of wave velocity, the square of
oscillation frequency, and the product of localization region size and
oscillation frequency, respectively. In the case $D_0 = 0$, we have the
Landau--Khalatnikov equation for the conserved ($D = 1$) and
nonconserved ($D = -1$) order parameters and the traveling-wave
equation ($D = 0$). The constant $c$ reduces to the diffusion coefficient
in the first case, to the inverse relaxation time in the second case,
and to the wave velocity in the third case. And finally, in the static
case $D_0 = -1$, we obtain the Poisson equation at $D = 1$, the
equilibrium condition at $D = -1$, and the Debye screening equation at
$D= 0$.
Here, $c$ reduces, respectively, to the square of screening radius, to
the inverse susceptibility ${\rm d}\mu/{\rm d}\eta$,
and to the screening radius.


It is evident that the approach outlined enables one to represent not
only linear equations of motion but also to generalize these to the
nonlinear case and to take into account the higher dispersion terms.
For example, if, by analogy with the theory of solitons \cite{9},
we set that, in addition to the
term $\propto k^D$, the flux involves the term $\propto k^{2+\Delta}$
with the exponent
$-1\le\Delta\le1$,
then a dispersion component of a higher order $3+\Delta$
appears. The nonlinearity is reflected in the dependence $\mu(\eta)$
of the chemical potential on the order parameter. One should take into
account that the dimension $\delta$ of a nonlinear channel does not
reduce to the corresponding quantity $D$ for a linear one. As a result,
equation (30) takes the form
\begin{eqnarray}
& &{{\rm d}^{1+D_0}\eta \over
{\rm d}t^{1+D_0}}+i^{1+D}c_0{{\rm d}^{1+D}\eta
\over {\rm d}x^{1+D}}+\beta {{\rm d}^{3+\Delta}\eta \over {\rm
d}x^{3+\Delta }}+{{\rm d}^{1+\delta}f(\eta )\over {\rm d}x^{1+\delta}}=0,
\nonumber\\
& &f(\eta )={{\rm e}^{\alpha_\delta
\delta}\left| \xi_\delta -1\right| ^\delta l_\delta^{1+\delta}\over
{\rm e}^{\alpha_0 D_0}\left| \xi_0-1\right| ^{D_0} \tau^{1+D_0}}
{{\rm d}\widetilde{\mu}\over {\rm d}\eta},
\end{eqnarray}
where $c_0$ corresponds to $\eta = 0$, $\beta > 0$ is the dispersion
parameter, and $\widetilde{\mu}(\eta )$
is the nonlinear contribution into the
chemical potential. This equation reduces to the Korteweg--de Vries
equation at $D_0 = D = \Delta=\delta=0$ and
$f\propto\eta^2$,
to the nonlinear Schr\"odinger one at $D_0=D=0$, $\Delta =\delta =-1$,  $f=\eta^3$,
and finally to the sine-Gordon
equation at $D_0=D=1$, $\beta =0$,  $\delta =-1$,  $f=\sin\eta$.
As is known, these equations are notable in that
their symmetry within the framework of the inverse problem of the
scattering theory
provides the existence of soliton solutions \cite{9}.
Here the question arises:
Will this symmetry also manifest itself at other values of the
differentiation orders $D$, $D_0$, $\Delta$, and $\delta$ and/or with
different choices of $\beta$ and $f(\eta)$? To answer this question
demands investigation of the symmetry of multidimensional fractal sets and, to
our knowledge, this has not been done.

Now, one can note only
that conformal symmetry (see \cite{10}) states the follow connection
between the above exponents:
\begin{equation}
D=D_0=2(1+\delta-n)^{-1}.
\end{equation}
Here exponent $n$ determines the nonliniar contribution
$f(\eta)\propto \eta^{n-1}$.
Within framework of the Landau $\eta^4$-theory, one has $n=4$
for a thermodynamic system characterised by the
additive noise which intensity is the temperature.
But with passing to the multiplicative noise with the amplitude
proportional to $\eta^a$, $0<a<1$ the powers of $\eta$ for all
terms in Eq.(31), excluding the first,
are lowered by the value $2a$ \cite{11}.


\newpage

\begin{thebibliography}{00}


\bibitem{1} A.I.~Olemskoi, {\it Fractals in Condensed Matter
Physics}, Ed. I.M.~Khalatnikov {\bf 18}, Part~1, Gordon \& Breach,
London, 1996.

\bibitem{2} R.R.~Nigmatullin, {\it Teor. Mat. Fiz.}, {\bf 90}, 354, 1992.

\bibitem{3} S.G.~Samko, A.A.~Kilbas, O.I.~Marichev,
{\it Integraly i Proizvodnye Drobnogo Poryadka i ikh Prilozheniya}
(Fractional-Order Integrals and Derivatives and their Applications)
Minsk, Nauka i Tekhnika, 1987.

\bibitem{4} D.~Forster {\it Hydrodynamical Fluctuations, Broken Symmetry,
and Correlation Functions}, Moscow, Atomizdat], 1980.

\bibitem{5} D.N.~Zubarev {\it Neravnovesnaya Statisticheskaya Termodinamika}
(Nonequilibrium Statistical Thermodynamics), Moscow, Nauka], 1971.

\bibitem{6} A.I.Olemskoi, I.V.Koplyk, Physics-Uspekhi, {\bf 38}, 1061, 1995.

\bibitem{7} A.I.~Olemskoi, A.Ya.~Flat,
{\it Usp. Fiz. Nauk}, {\bf 163}, No. 12, 1, 1993.

\bibitem{8} A.I.~Olemskoi, A.Ya.~Flat,
{\it Sumy: Vestnik SumGU}, {\bf 1}, 10, 1994.

\bibitem{9} V.E.~Zakharov, S.F.~Manakov, S.P.~Novikov,
and L.P.~Pitaevskii {\it Teoriya Solitonov: Metod Obratnoi Zadachi}
(Soliton Theory: Inverse Problem Approach), Moscow, Nauka, 1980.

\bibitem{10} N.Kh.~Ibragimov {\it Gruppy preobrazovaniy
v matematicheskoy fizike}
(Transformations Groups in Mathematical Physics),
Moscow, Fizmatgiz, 1983.

\bibitem{11} A.I. Olemskoi, Physics--Uspekhi, {\bf 41}, 269, 1998.

\end{thebibliography}
\end{document}